\documentclass[preprint,12pt,authoryear]{elsarticle}

\usepackage{threeparttable}
\usepackage{array}
\usepackage{gensymb}
\usepackage{siunitx}
\usepackage{tabu}
\usepackage{graphicx}
\usepackage{listings}
\usepackage{natbib}

\usepackage[hidelinks]{hyperref}
\hypersetup{breaklinks=true}
\journal{arXiv}
\usepackage{amssymb}

\begin{document}
\begin{frontmatter}

\title{Combined Location Online Weather Data: Easy-to-use Targeted Weather Analysis for Agriculture}

\author{Darren Yates$^a$$^e$, Christopher Blanchard$^b$, Allister Clarke$^b$, Sabih-Ur Rehman$^a$, Md Zahidul Islam$^a$, Russell Ford$^c$, Rob Walsh$^d$ }

\address{$^a$School of Computing, Mathematics and Engineering, Charles Sturt University, Bathurst NSW 2795, Australia\\
Email: \{dyates, sarehman, zislam\}@csu.edu.au\\
$^e$corresponding author}
\address{$^b$Gulbali Institute, Charles Sturt University, Wagga Wagga NSW 2640, Australia\\
Email: \{cblanchard, aclarke\}@csu.edu.au\\
}
\address{$^c$RLF Agro R\&D Consulting, New South Wales, Australia\\
Email: rusty58ford@gmail.com\\
}
\address{$^d$Ricegrowers Ltd (SunRice), Leeton, NSW, Australia\\
Email: rwalsh@sunrice.com.au\\
}

\begin{abstract}
The continuing effects of climate change require farmers and growers to have greater understanding of how these changes affect crop production. However, while climatic data is generally available to help provide much of that understanding, it can often be in a form not easy to digest. The proposed Combined Location Online Weather Data (CLOWD) framework is an easy-to-use online platform for analysing recent and historical weather data of any location within Australia at the click of a map. CLOWD requires no programming skills and operates in any HTML5 web browser on PC and mobile devices. It enables comparison between current and previous growing seasons over a range of environmental parameters, and can create a plain-English PDF report for offline use, using natural language generation (NLG). This paper details the platform, the design decisions taken and outlines how farmers and growers can use CLOWD to better understand current growing conditions. Prototypes of CLOWD are now online for PCs and smartphones. 
\end{abstract}


\end{frontmatter}

\section{Introduction}
\label{}
The collection and analysis of weather data has long been a fundamental part of the agriculture industry \citep{eca001,eca025}. However, as the effects of climate change continue to play out across the globe, data analysis is playing an ever-increasing role in understanding those effects. In Australia, daily weather data is continuously recorded through a nation-wide series of weather stations under the guidance of the Australian Bureau of Meteorology. This data is collated each day by the SILO database project, hosted by the Queensland Government Department of Environment and Science\footnote{SILO project - \url{https://www.longpaddock.qld.gov.au/silo/}} \citep{eca010}. For a number of weather station locations, this SILO data is available dating back as far as 1899.\\
\indent The SILO project provides access to this raw data through its website. However, there is opportunity to deliver more targeted data analysis, particularly for agricultural purposes, as well as easier access on mobile devices. This paper introduces the proposed Combined Location Online Weather Data (CLOWD) framework that focuses on these goals. CLOWD enables users to choose the location and start date for data analysis, thereby enabling a focus on local growing seasons. It also locally-computes additional data, including growing-degrees days (GDD) \citep{eca003} and incorporates natural language generation (NLG) \citep{eca009} technology to auto-generate plain-language PDF reports. \\
\indent Moreover, CLOWD requires only a standard (HTML5) web browser and an internet connection, thus, no additional applications need to be downloaded and installed on the user's device. Prototypes of the CLOWD platform are now available for PCs/laptops\footnote{CLOWD for PCs/laptops - \url{https://clowd.csu.edu.au}} and smartphones/tablets\footnote{CLOWD for smartphones/tablets - \url{https://clowds.csu.edu.au}}.

\begin{table}
\centering
\footnotesize
\renewcommand{\arraystretch}{1.2}
\caption{Table of definitions for additional acronyms used in this paper}
\begin{tabular}{llll}
\hline
API&Application Programming Interface&NLG&Natural Language Generation\\
CSS&Cascading Style Sheet&NSW&New South Wales\\
DST&Decision Support Tool&PDF&Portable Document Format\\
GDD&Growing Degree Days&UI&User Interface\\
HCI&Human-Computer Interaction&UX&User eXperience\\
HTML&HyperText Markup Language&VPD&Vapour-Pressure Deficit\\
\hline
\end{tabular}
\label{table:acc}
\end{table}

\subsection{Original contributions}
As will be detailed here, the original contributions of this paper include:
\begin{itemize}
\item A framework that enables comparative agriculture-focused weather data analysis to be performed on historical and recent Australian data using a web browser on either a PC or smartphone.
\item A framework that returns detailed weather analysis of any location in Australia with a single map-click.
\item A framework that starts the growing season on any user-selected date in the year (day-zero) and initialises all data analysis to that date.
\item A framework that incorporates natural language generation (NLG) methods to help simplify understanding of chart trends.
\end{itemize}
\indent While acronyms used frequently in this paper are listed in Table \ref{table:acc}, this paper will now continue with a look at related research in Section 2. This will be followed by details of the CLOWD framework in Section 3 and additional locally-processed data analysis functions in Section 4. An example of discovered knowledge with reference to the Australian rice industry will follow in Section 5, then discussion regarding future research opportunities in Section 6. Finally, section 7 will conclude this research paper.

\section{Related work}
\label{sec:relres}
Decision support tools (DSTs) are an increasingly important component in digital farming around the world \citep{eca018}, with those that incorporate smart systems, big data and artificial intelligence, considered part of `Agriculture 4.0' \citep{eca017}. However, research definitions of DSTs and the outcomes they provide are known to vary. A DST has been previously defined as `a human-computer system which utilises data from various sources, aiming at providing farmers with a list of advice for supporting their decision-making under different circumstances', but one that does not go so far as to provide direct instructions \citep{eca017}. However, other definitions are more broad, with agriculture-focused DSTs categorised alternatively as either `generating a series of evidenced-based recommendations', or acting more as information systems, leaving the end user (in this case, a farmer) to come to their own conclusions \citep{eca018}.\\
\indent Nevertheless, research suggests the use of DSTs over the last ten years has by no-means been universal. A research survey in 2016 revealed that the use of DSTs in both the U.K. and other regions was `limited' \citep{eca018}. This result was further supported by research conducted in 2019 by Michels, Bonke and Musshoff \citep{eca002}. This research identified a very high level of technology adoption from a survey of 207 German farmers, with 95\% of those surveyed using a smartphone, yet found that adoption of phone-based crop-protection DSTs was only 71\% \citep{eca002}.\\
\indent Moreover, the research by Michels et al attempted to identify the reasons why farmers in the survey chose or chose not to use the phone-based crop-protection DSTs and made some noteworthy findings. First, the analysis of farmer responses was applied to the Unified Theory of Acceptance and Use of Technology, which considers four key behavioural factors that include performance expectancy, effort expectancy, social norm and facilitating conditions \citep{eca019}. In summary, performance expectancy is the level of improvement a user believes an app has made to their task performance, while effort expectancy is essentially the `ease of use' a person feels using an app. Social norm is the level a user feels they should be expected to be using such a app, while facilitating conditions is the user's perception of how well existing technology supports the app's key functions \citep{eca002,eca019}.\\
\indent Given a survey asking which functions were considered the most useful in apps related to crop-protection, 77\% of 198 respondants chose `weather information' and 83\% said they were already using it. The authors made the caveat that many smartphones already include weather information, but did not specify what type of weather information was used by farmers and whether a built-in app featured such information. Nevertheless, weather information was perceived by respondents of the survey as the most useful of features \citep{eca002}. The second point was in considering facilitating conditions, the authors noted the importance of the smartphone being capable of installing the app \citep{eca002}. For example, an iOS (Apple) app will not install on Android devices and vice versa. Importantly, our proposed CLOWD framework bypasses this issue by being a web-based app that requires no additional software installation.\\
\indent The third point from this research was the importance of a DST (in particular, the user interface) being easy to use \citep{eca002}. This point is supported also by \citet{eca017} where one of the key challenges in the development of `Agriculture 4.0' DSTs was simplifying the DST's graphical user interface. It is in light of these findings that our proposed CLOWD framework exists as two separate platforms - one for smartphones and one for PCs/laptops, each with a completely separate and targeted user interface. This paper will now discuss this framework in the following section.

\section{CLOWD development framework}
\begin{figure}
\centering
\includegraphics[width=0.9\columnwidth]{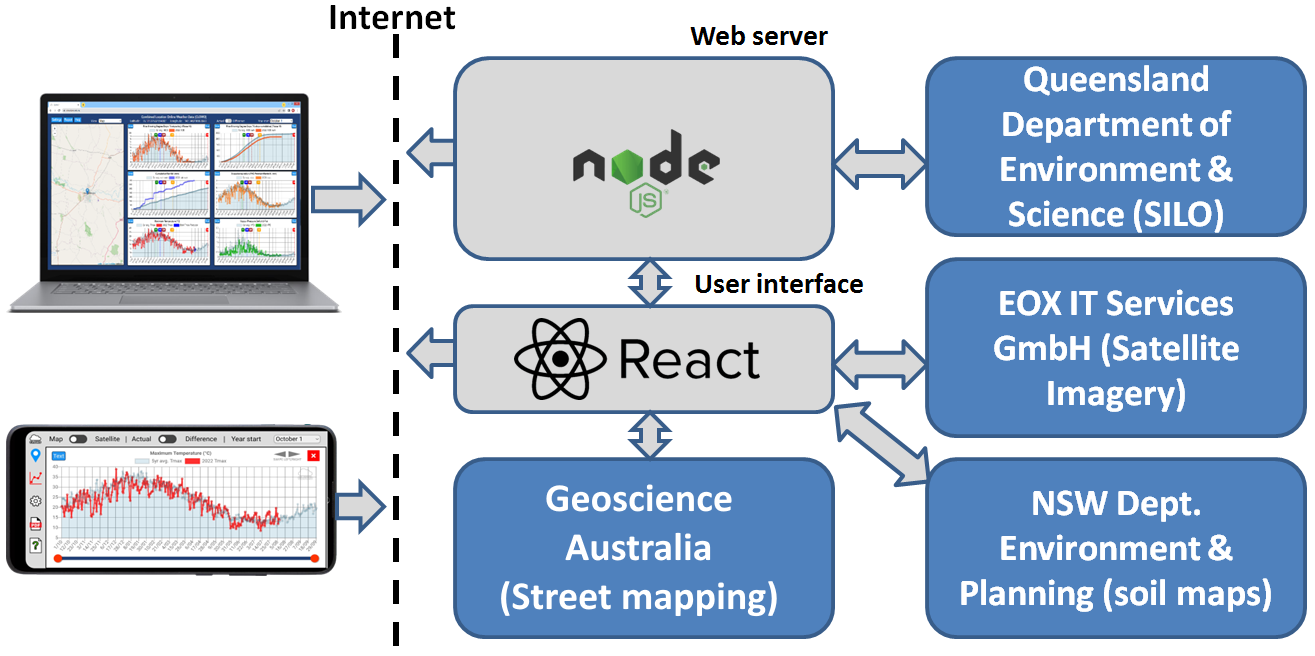}
\caption{The CLOWD framework features the Node.js JavaScript web service software that connects to the Queensland Department of  Environment and Science's SILO historical weather data server, with the React user interface development system used to directly access location data from Geoscience Australia, EOX IT Services GmbH's Sentinel-2 satellite imagery and soil data from the NSW Department of Environment and Planning.}
\label{fig:clowd1}
\end{figure}

The proposed CLOWD framework, shown in Fig. \ref{fig:clowd1}, is divided into two sections. First, the user interface or `front end' controls delivery of information to the user via their web browser through the React open-source user interface platform. It also directly accesses location and soil map data from various CC-BY-4.0-licensed sources discussed in Section \ref{sec:datsrc}. Second, the web server-side or `back end' section interacts with, as well as collects the required data from, the Queensland Department of Environment and Science `SILO' server. This section is built on the Node.js open-source JavaScript server platform.  In all cases, data is retrieved through open-access application programming interfaces (APIs). This chapter will now describe these sections in more detail, including decisions taken for various design elements.

\begin{figure}
\centering
\includegraphics[width=1\columnwidth]{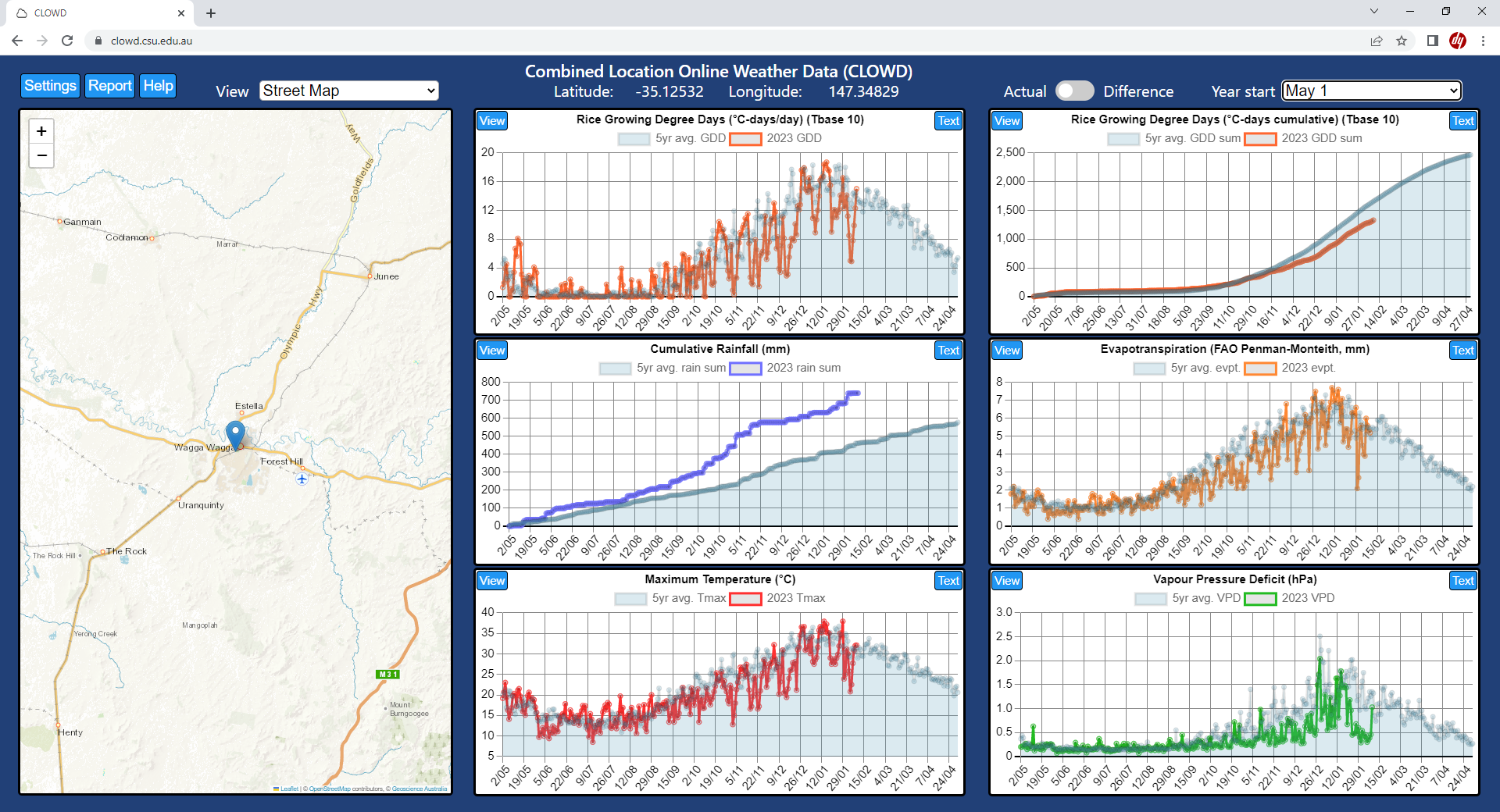}
\caption{The desktop version of CLOWD gives the user the ability to analyse multiple weather data for any location in Australia at the click of a map.}
\label{fig:clowd2}
\end{figure}

\begin{figure}
\centering
\includegraphics[width=1\columnwidth]{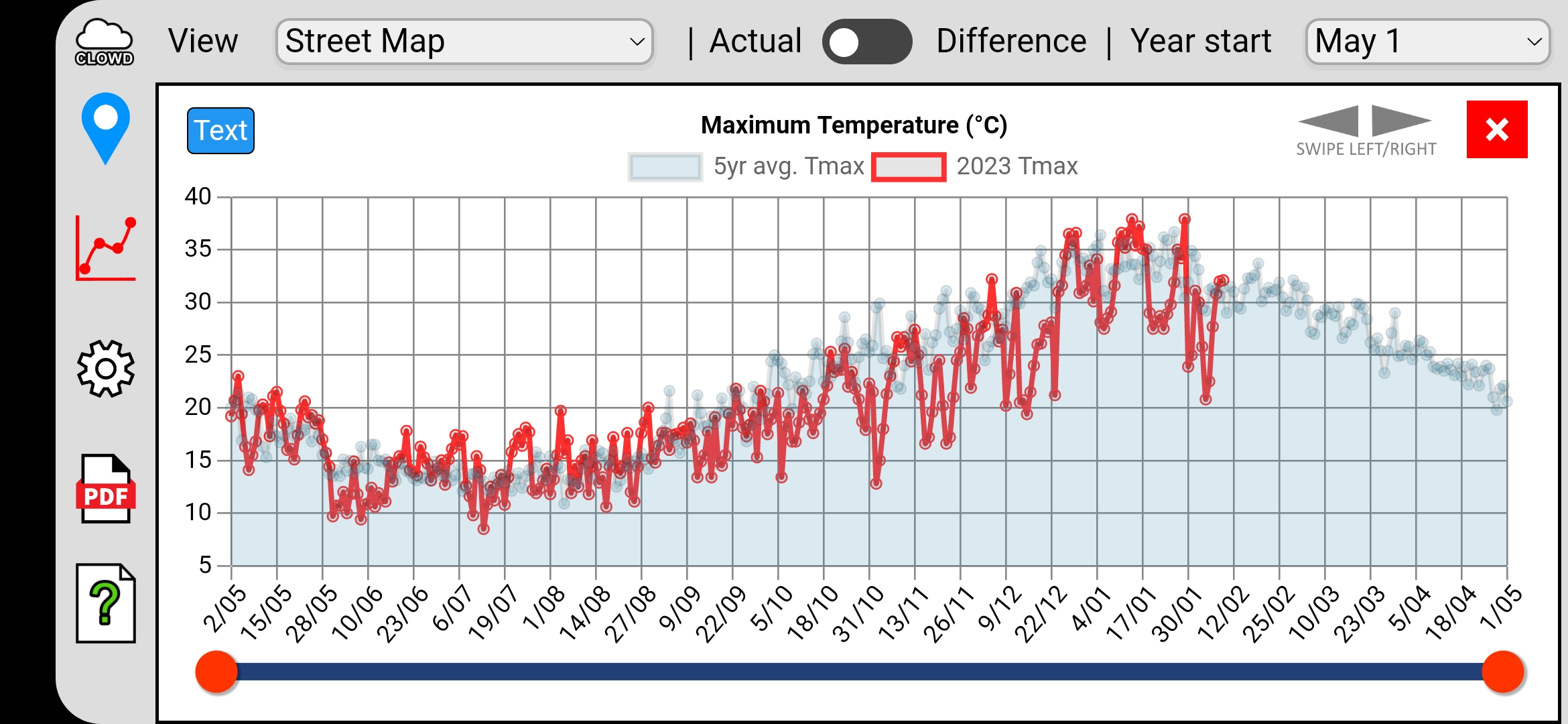}
\caption{The many data charts available in the mobile version of CLOWD are selected by swiping either left or right on the chart screen. The chart data can be magnified using the slide-magnifier control and individual chart traces can be viewed by tapping the coloured legend tiles.}
\label{fig:clowd3}
\end{figure}

\subsection{The User Interface Design}
The user interface design centres on the user selecting a location anywhere within Australia on a zoomable/clickable map and viewing the  analysis of weather data as near to that location as data is available through a series of charts, each chart featuring one weather attribute. Thus, the goal is to keep the user interface as simple as possible to facilitate ease and speed of use.\\
\indent Moreover, to better facilitate its use on different screen sizes, two versions of the user interface have been developed - a six-charted version for large/high-resolution PC and laptop screens as shown in Fig. \ref{fig:clowd2}, along with a second single-chart screen version for smartphones, shown in Fig. \ref{fig:clowd3}. Thus, following research in Section \ref{sec:relres}, the purpose of designing separate user interfaces was to make use of the inherent advantages of each platform. For example, larger-dimensioned PC/laptop screens enable the use of six side-by-side charts to more quickly determine trends between weather attributes. The single-charted phone version incorporates left/right-swipe control for fast access between the various chart options on smaller screens, whilst maintaining readability.\\
\indent These design choices were also made on the basis of previous research in human-computer interaction (HCI) that identified the relationship between functional complexity and interaction simplicity required in modern UI design \citep{eca006,eca007}. The methodology used in designing both interfaces is to maintain what is considered as `internal consistency', which is the consistency of the interface across the application components. This can include consistency in placement and design, ensuring the user has a simple user-experience (UX). Research by \citet{eca008} has more recently shown that usability (how easy a website is to use) is affected by readability (how easy it is to read). They identified a number of useful guidelines for improving the overall experience, including left-justified text, using font sizes greater than 12-point and the use of sans-serif style fonts. Where possible, these design guidelines have been implemented in the CLOWD user interface designs, to further reduce any barriers for users to benefit from these web applications.\\
\indent In both versions, the charts are customisable, in terms of the data and weather attributes they show. Moreover, the map allows a choice of a street view, satellite image, soil type (NSW only) or estimated soil fertility (NSW only) of the area selected. All map options are coordinated such that they remain in lock-step as the map is moved and scaled.\\
\indent The desire for a smooth and easy UX also led to the inclusion of a PDF (Portable Document File format) help/instruction manual to detail the CLOWD functionality. This appears inside a local PDF viewer on both versions. The importance of this delivery method is that it ensures the user always has the latest reference help document available as it is automatically downloaded and displayed upon pressing the `Help' button. 

\begin{figure}
\centering
\includegraphics[width=1\columnwidth]{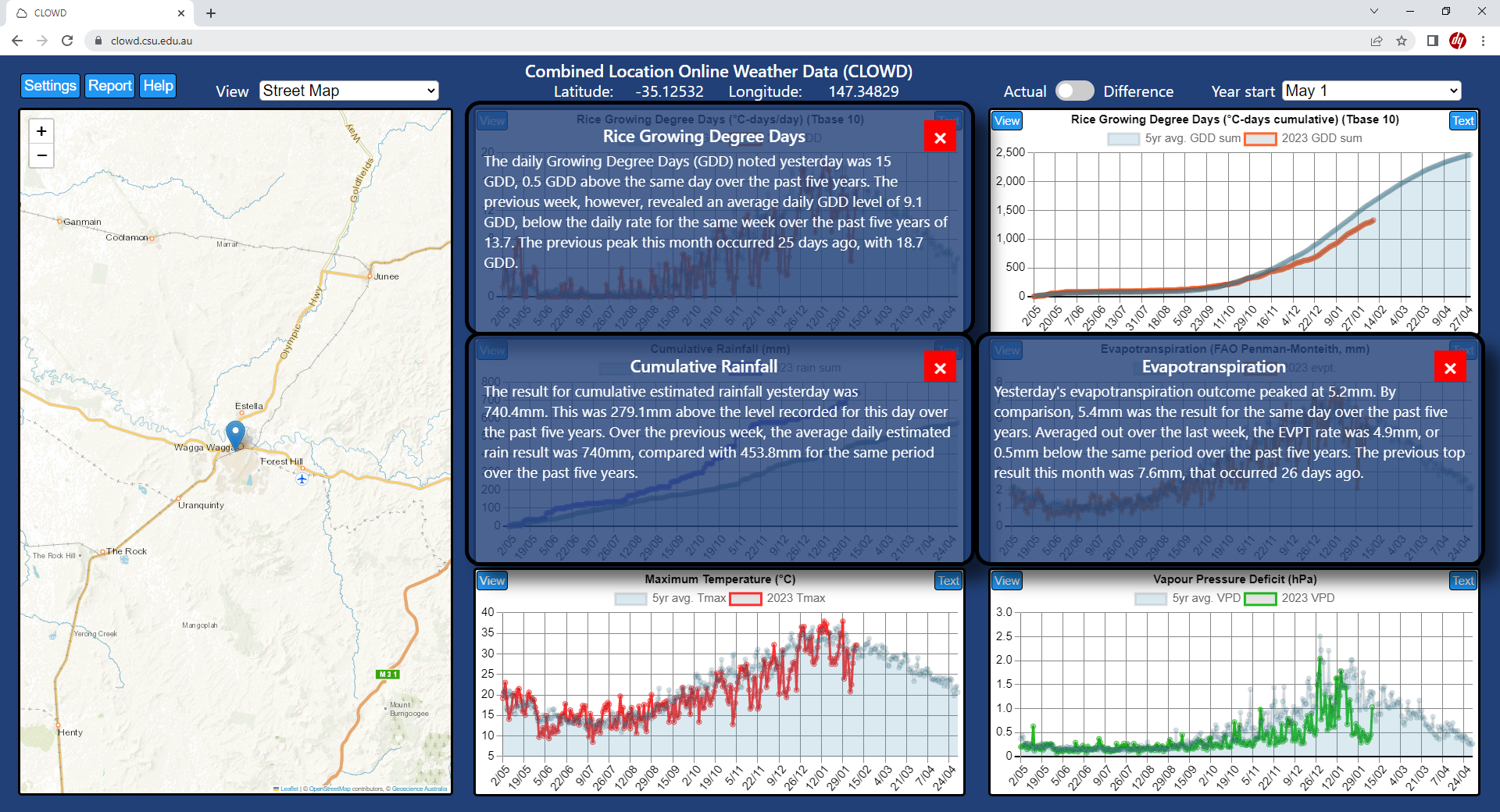}
\caption{Text used in the PDF report can be seen on individual data charts through each chart's Text button and is created automatically via Natural Language Generation (NLG). }
\label{fig:clowd4}
\end{figure}

\subsubsection{Natural Language Generation (NLG) text}
Research has long shown that understanding graphs and charts has been an important part of basic mathematical learning \citep{eca020}, but also an area where users may struggle with understanding \citep{eca021}. To help mitigate this, each CLOWD chart incorporates a `text' version automatically created through a programming technique called `natural language generation' (NLG) \citep{eca009}. In its simplest form, NLG contains a series of pre-coded natural-language sentences that are then populated with the appropriate data and delivered to the user's screen. More complex forms automatically alter the sentence structures based on the data values. The purpose of NLG in CLOWD is to provide an overview of the chart's key points and trends in a text form that can be read easily by those who may struggle with understanding charts. The NLG functionality is coded into the user interface and does not require the CLOWD server to process it, helping to lower server costs. This feature can be seen in Fig. \ref{fig:clowd4}.
\subsubsection{Auto-generated PDF report}
\begin{figure}
\centering
\includegraphics[width=1\columnwidth]{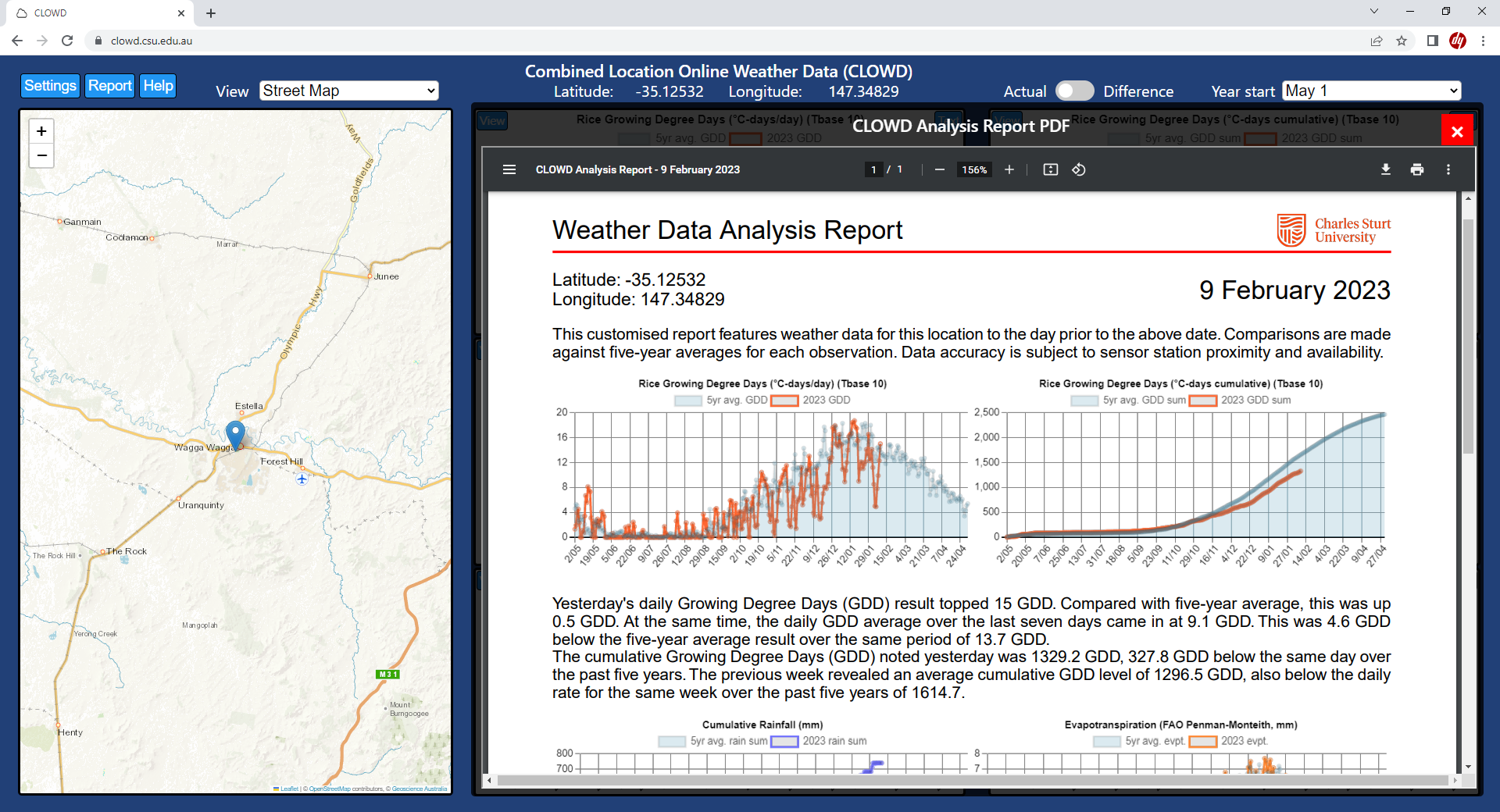}
\caption{CLOWD can auto-generate a single-page PDF report featuring the current-location data charts and text summary using natural language generation (NLG).}
\label{fig:clowd5}
\end{figure}

\begin{figure}
\centering
\includegraphics[width=1\columnwidth]{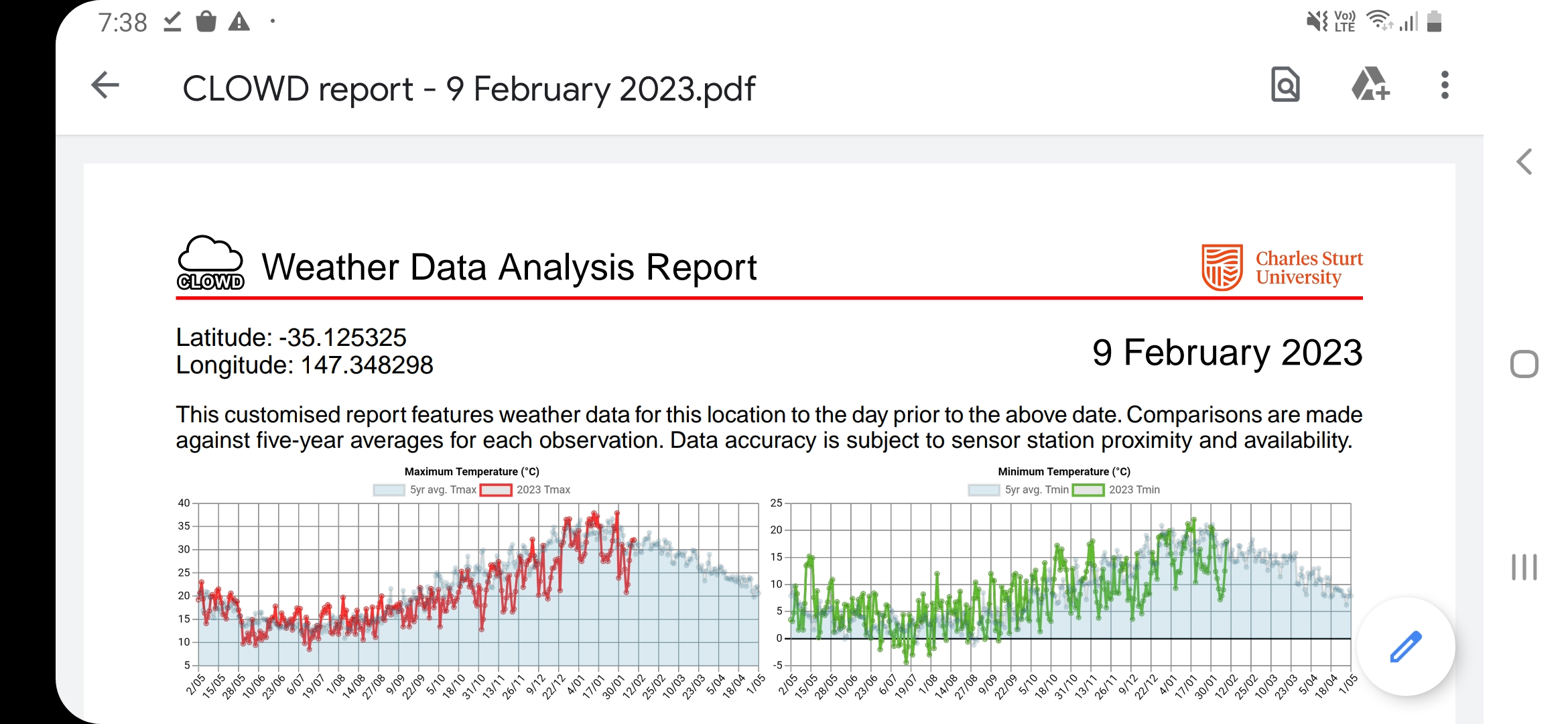}
\caption{A PDF report can be generated from selected data charts within the CLOWD web application directly on the mobile device.}
\label{fig:clowd6}
\end{figure}

\begin{figure}
\centering
\includegraphics[width=1\columnwidth]{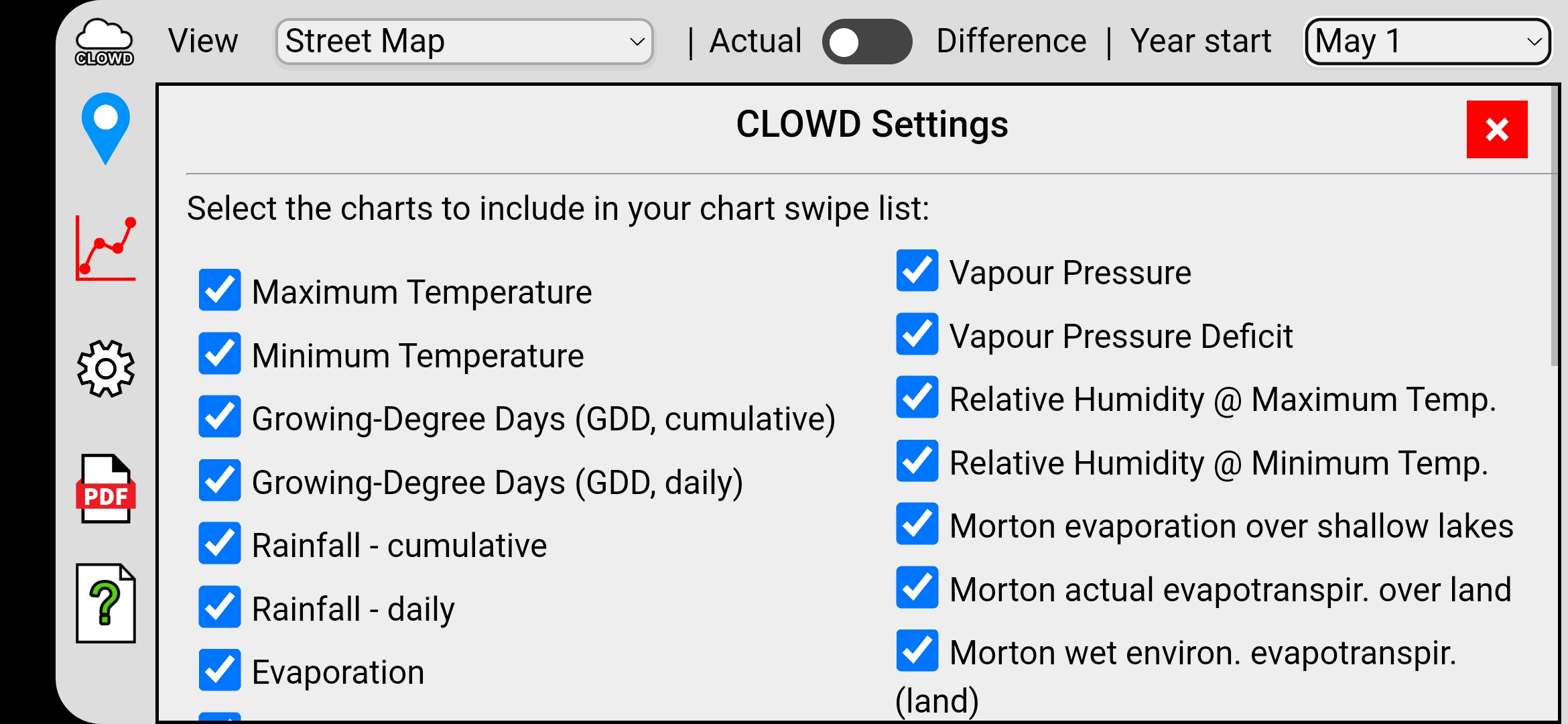}
\caption{Up to 18 data charts are available and are selected for viewing rotation through the settings button on the left-side menu.}
\label{fig:clowd7}
\end{figure}

The same NLG techniques are also combined to enable the creation of a location-specific weather analysis report. This PDF creation again takes place on the client device, relieving any pressure on the CLOWD web server to handle the load. On the PC/laptop version of CLOWD, the single-page PDF features all six visible selected charts in addition to their NLG text, as can be seen in Fig. \ref{fig:clowd5}, and can be viewed within the CLOWD application user interface. However, the PDF report can also be downloaded to the client device's local storage for further use or dissemination.\\
\indent The smartphone version of the PDF report is similarly generated on the smartphone itself and is shown in Fig. \ref{fig:clowd6}. However, the choice of charts and accompanying NLG text is determined by the chart rotation list created by the user within the CLOWD settings page, as shown in Fig. \ref{fig:clowd7}. All 18 possible charts can be included in the smartphone PDF report, which is created with a similar `2x3'-layout (two columns of three charts per column) per page and automatically expands to include all selected charts. 

\subsection{User Interface technology framework}
The user interface is developed using the React framework, originally developed by Facebook (now Meta) and made open-source in 2016 \citep{eca004}. React describes itself as a `JavaScript library for building user interfaces'. It simplifies the task of interacting between data manipulation processes within the JavaScript programming language and linking the outcomes of those processes with on-screen controls and displays. As the user interface appears within a standard web browser, this is still coded in standard HTML (HyperText Markup Language) and CSS (Cascading Style Sheets) languages. However, the additional analysis applied to the retrieved weather data in CLOWD is a combination of Node.js (retrieval), React (processing) and custom JavaScript functions developed for CLOWD, a number of which will be detailed in Section \ref{sec:ana}.\\
\indent Numerous additional libraries available via the Node.js Package Manager (NPM) are utilised within the React-powered user interface, including the Leaflet and React-leaflet map retrival/display modules.\\
\indent On-device analysis reduces the processing load on the CLOWD server, but does not appear to have a noticeably detrimental effect on client device performance, including recent smartphones. Previous research has shown that smartphones are fast approaching - and even exceeding - the processing performance levels of current-generation laptops \citep{eca005}.

\subsection{The web server design}
As is shown in Fig. \ref{fig:clowd1}, the web server provides a dual role in delivering the React user interface to the user's web browser, but also collecting the required data from third-party providers in response. This includes the selected-location weather attribute data. However, because attribute data analysis takes place on the user/client device, the server does not require as high a level of application processing performance. Thus, depending on the number of concurrent users, the level of performance required by the server can be expected to be more modest, again helping to lower system costs.\\
\indent The CLOWD web server is developed using the Node.js open-source server environment. Node.js is cross-platform, enabling it to run on Windows, Linux and macOS computing platforms. Thus, the CLOWD framework is also cross-platform, with the prototype tested on an Amazon Web Services Linux-based server. Node.js incorporates a library of third-party add-ons to provide additional functionality, such as the `react-chartjs-2'\footnote{react-chartjs-2 - \url{https://react-chartjs-2.js.org/}} charting library, plus the Leaflet and React-leaflet\footnote{React Leaflet - \url{https://react-leaflet.js.org/}} libraries mentioned earlier.

\subsection{Data Sources}
\label{sec:datsrc}
\begin{figure}
\centering
\includegraphics[width=1\columnwidth]{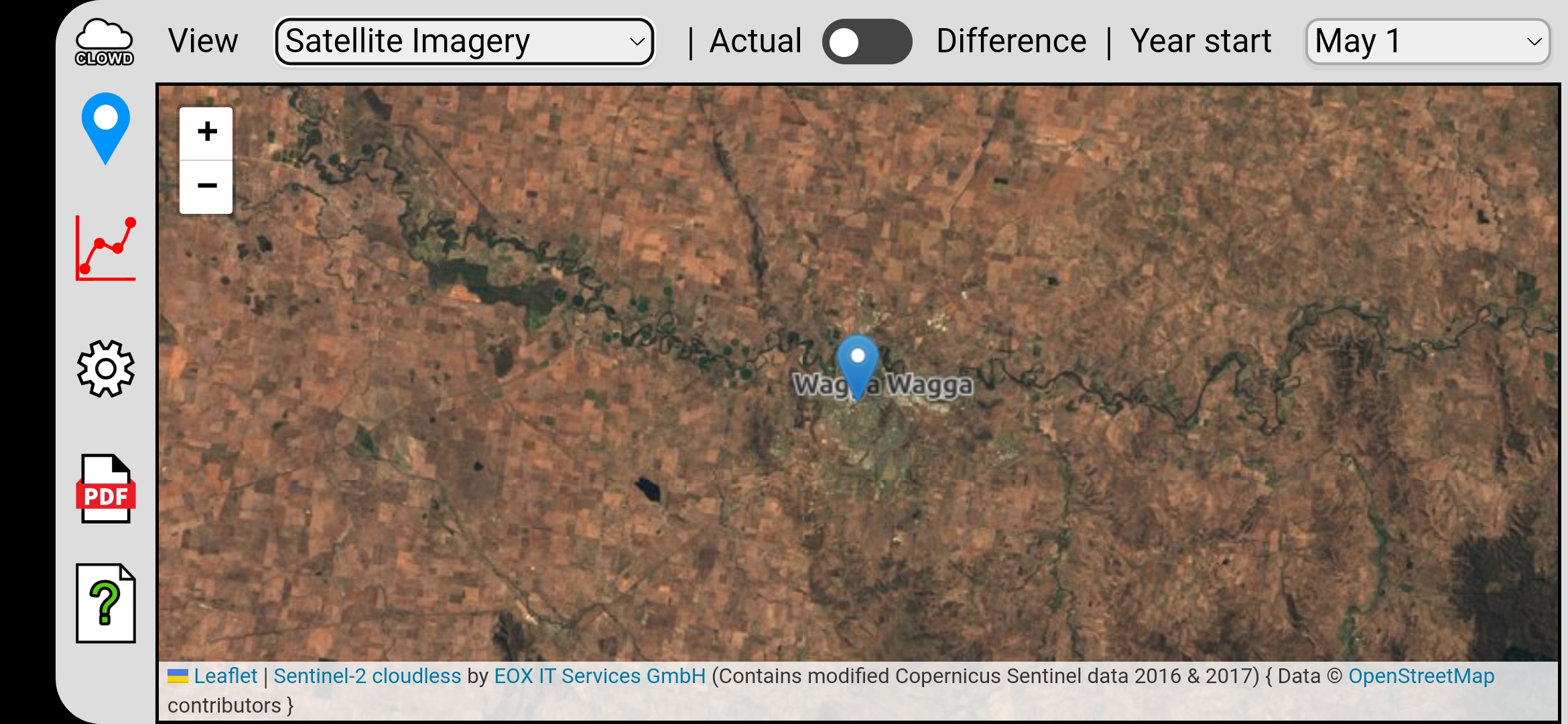}
\caption{The mobile version includes the same functionality reduced for phone-screen consumption, including lock-step map and satellite location selection and supports both Android and iOS devices.}
\label{fig:clowd8}
\end{figure}

The key source of weather data for CLOWD is the SILO project\footnote{Queensland DES SILO - \url{https://www.longpaddock.qld.gov.au/silo/}}, produced by the Queensland Government Department of Environment and Science and offers over 120 years of daily meteorological data for locations around Australia. The data is made available free-of-charge via a Creative Commons Attribution 4.0 International (CC-BY-4.0) license through an application programming interface (API).\\
\indent Street map data is provided by Geoscience Australia\footnote{Geoscience Australia - \url{https://ga.gov.au/}} through a Creative Commons Attribution 4.0 license and incorporated into CLOWD via the Leaflet JavaScript module. Satellite imagery is sourced from the `Sentinel-2 cloudless 2016' data captured during 2016/17 and provided under a Creative Commons 4.0 international licence by EOX IT Services GmbH\footnote{EOX IT Services GmbH - \url{https://s2maps.eu/}}. Although not specified, the resolution of the satellite imagery is sufficient for recognition down to paddock level - in combination with the Geoscience Australia street view, this is all CLOWD requires. The imagery is provided through APIs, allowing images to appear in lock-step (both position and scale) with the Geoscience Australia map view. Fig. \ref{fig:clowd8} shows this through the `View' dropdown-box. Thus, data used within the CLOWD framework is either freely available or licensed via attribution within the CLOWD user interface.\\
\indent The following section will now describe some of the additional analysis functions provided within the CLOWD framework.

\section{Data analysis}
\label{sec:ana}
While the CLOWD framework utilises raw weather data from the SILO project, it also provides additional analysis of important agricultural parameters that are calculated locally on the client device. These include growing-degree days (GDD) and vapour-pressure deficit (VPD), along with analysis of recent temperatures (maximum and minimum) to provide targeted alerts.
\subsection{Growing-degree days (GDD)}
The biological processes of growing crops rely on heat and the mathematical measure of `growing-degree days' (GDD) is commonly used to indicate the timing of these processes or phenology stages \citep{eca003}. GDD is generally calculated daily and a crop accumulates GDD to provide an indication of the likely current phenology or growth stage. While the SILO project does not calculate GDD, it provides the minimum and maximum temperatures and can be calculated on the client device. The daily equation takes the form:
\begin{equation}
GDD = [(T_{max} + T_{min})/2] - T_{base}
\end{equation}
where $T_{max}$ is the maximum temperature over the 24 hour period, $T_{min}$ the minimum temperature and $T_{base}$ the crop-specific threshold temperature \citep{eca003}. However, as the work by \citet{eca003} examines, there are two interpretations for how GDD is calculated, with regards to when the daily average temperature is below the base temperature, $T_{base}$. The first is that if the result of the GDD calculation is less than zero, it is converted to zero (that is, it can never be negative). The second is if either $T_{max}$ or $T_{min}$ are less than $T_{base}$, these individual components are set to $T_{base}$. At present, the CLOWD framework implements the first interpretation.\\
While daily-GDD data is useful, it is commonly considered a cumulative measure, thus it is calculated as:
\begin{equation}
GDD_{sum} = \sum_{n=1} ^{k} GDD_{(n)}
\end{equation}
where $GDD_{(n)}$ is the GDD measure for the $n^{th}$ day in the growing season of $k$ days, such that $n={[1..k]}$. Phenology stages can be tracked based on the crop achieving a particular level of cumulative GDD, with the count beginning on the sowing date (the date CLOWD users would set as start or zero-day).

\subsection{Vapour-pressure deficit}
\label{sec:vpd}
The amount of water needed on a daily basis by a growing crop is dependent on the environmental vapour-pressure deficit (VPD) \citep{eca014}. Similarly to GDD, CLOWD calculates VPD using a series of equations \citep{eca022}, based on raw data provided by the SILO project. \\
\indent The first step is to calculate the average saturation vapour pressure, $e_{s(avg)}$. This is done by calculating the saturation vapour pressure at both $T_{min}$ and $T_{max}$ via the equations:
\begin{equation}
e_{s(T_{min})} = 0.6108exp(\frac{17.27T_{min}}{T_{min}+237.3})  
\end{equation}
\begin{equation}
e_{s(T_{max})} = 0.6108exp(\frac{17.27T_{max}}{T_{max}+237.3})  
\end{equation}
The two values are then averaged:
\begin{equation}
e_{s(avg)} = \frac{e_{s(T_{max})} + e_{s(T_{min})}}{2}  
\end{equation}
At this point, the actual vapour pressure (avp) is derived as a function of relative humidity:
\begin{equation}
avp=\frac{[e_{s(T_{min})}\times(RH_{max}/100)] + [e_{s(T_{max})}\times(RH_{min}/100)]}{2}  
\end{equation}
where $RH_{max}$ is the maximum relative humidity and $RH_{min}$, the minimum relative humidity, from SILO data. The assumption is also made here that the point of maximum vapour pressure occurs at the time of minimum relative humidity and vice versa \citep{eca015} (SILO records relative humidity as percentage values, hence these values are divided by 100 to obtain a proportional factor).\\
\indent Finally, the vapour pressure deficit (VPD) is the difference between the average saturated vapour pressure and the actual vapour pressure:
\begin{equation}
VPD = e_{s(avg)} - avp
\end{equation}
The CLOWD framework calculates VPD on-demand for each day on the client/user device before it is charted as required. A JavaScript function called `calculateVpd' has been created and incorporates the above equations to calculate VPD as follows:

\scriptsize
\begin{lstlisting}
calculateVpd(tmax, tmin, rhmaxt, rhmint) {
   var svpMax = 0.6108 * Math.exp( (17.27 * tmax) / (tmax + 237.3) );
   var svpMin = 0.6108 * Math.exp( (17.27 * tmin) / (tmin + 237.3) );
   var avgSvp = (svpMax + svpMin)/2.0;
   var avp = ( (svpMin * (rhmaxt/100)) + (svpMax * (rhmint/100)) ) / 2.0;
   return avgSvp - avp;
}
\end{lstlisting}
\normalsize

This function takes in the parameters $T_{max}$, $T_{min}$, $RH_{minT}$ and $RH_{maxT}$ retrieved from the SILO data and returns the calculated VPD value.

\subsection{Weather analysis alerts}
\begin{figure}
\centering
\includegraphics[width=1\columnwidth]{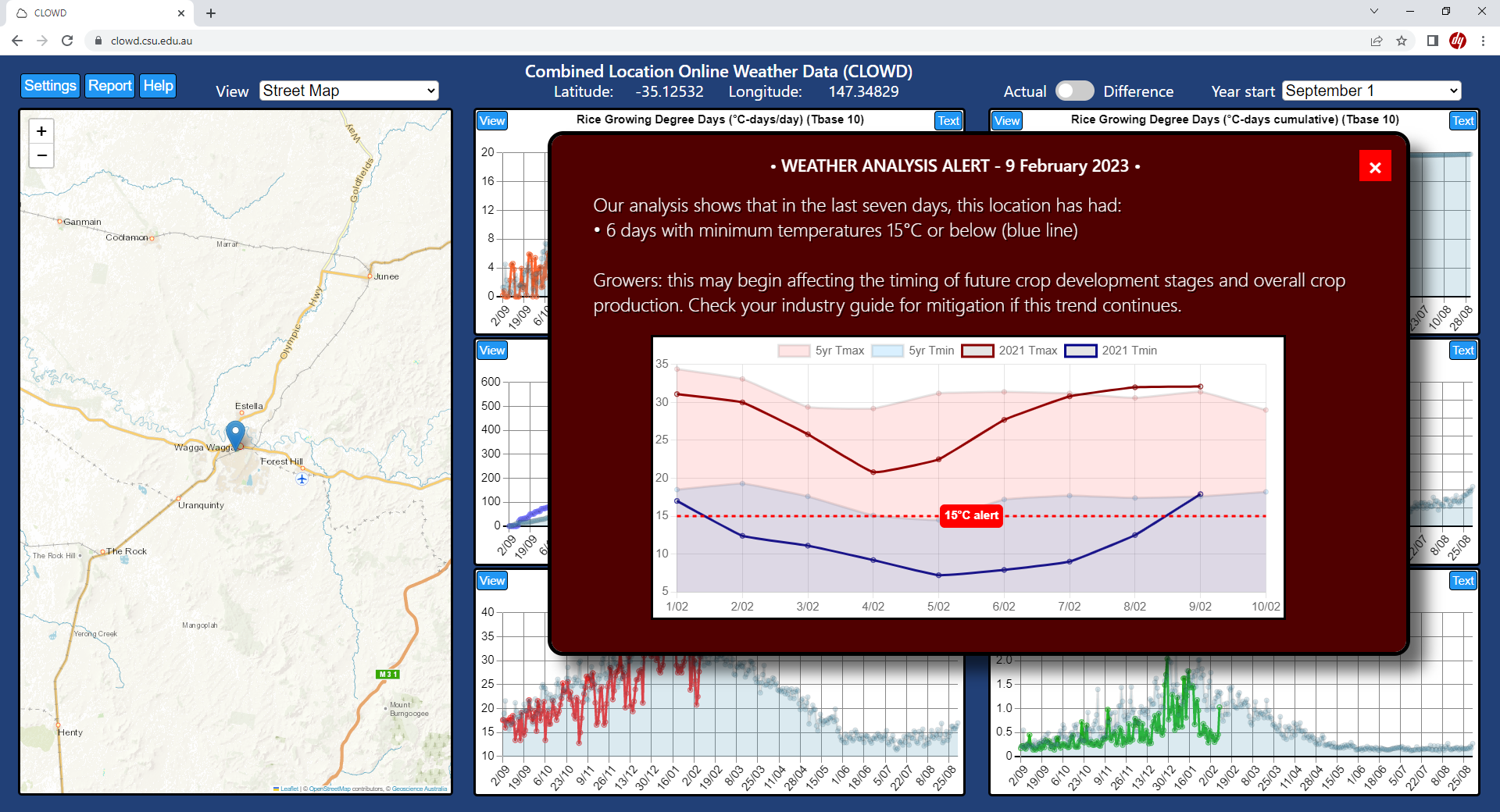}
\caption{Weather alerts are automatically generated when the maximum or minimum temperature exceeds the threshold set by the user in the CLOWD settings. This alert can also be deactivated as required by the user.}
\label{fig:clowd9}
\end{figure}

Weather conditions can have a detrimental effect on crop quality and yield. In Australia, it is generally recognised that night-time temperatures below 15$^{\circ}$C can cause spikelet sterility during the microspore phenology stage in varieties of rice \citep{eca016} and affect head rice yield.  However, this night-time temperature threshold varies not just between rice varieties but crop varieties also \citep{eca016}. Thus, a method that delivers an automated and customisable alert to crop growers could prove valuable.\\
\indent The CLOWD framework provides this alert system and is selectable by the user within the CLOWD settings page. An example is shown in Fig. \ref{fig:clowd9}. As the user selects a new location and in addition to the weather data for that location being downloaded from the SILO server, the preceding nine days are analysed for maximum and minimum temperature. If these temperatures are outside the threshold selected by the user (also chosen in the settings page), an alert pop-up window appears to notify the user that recent temperatures are outside the threshold and that additional crop care may be required. The threshold temperature setting is stored locally within the client device's web browser and activated as chosen by the alert settings option.

\section{Knowledge Discovery}
\begin{figure}
\centering
\includegraphics[width=1\columnwidth]{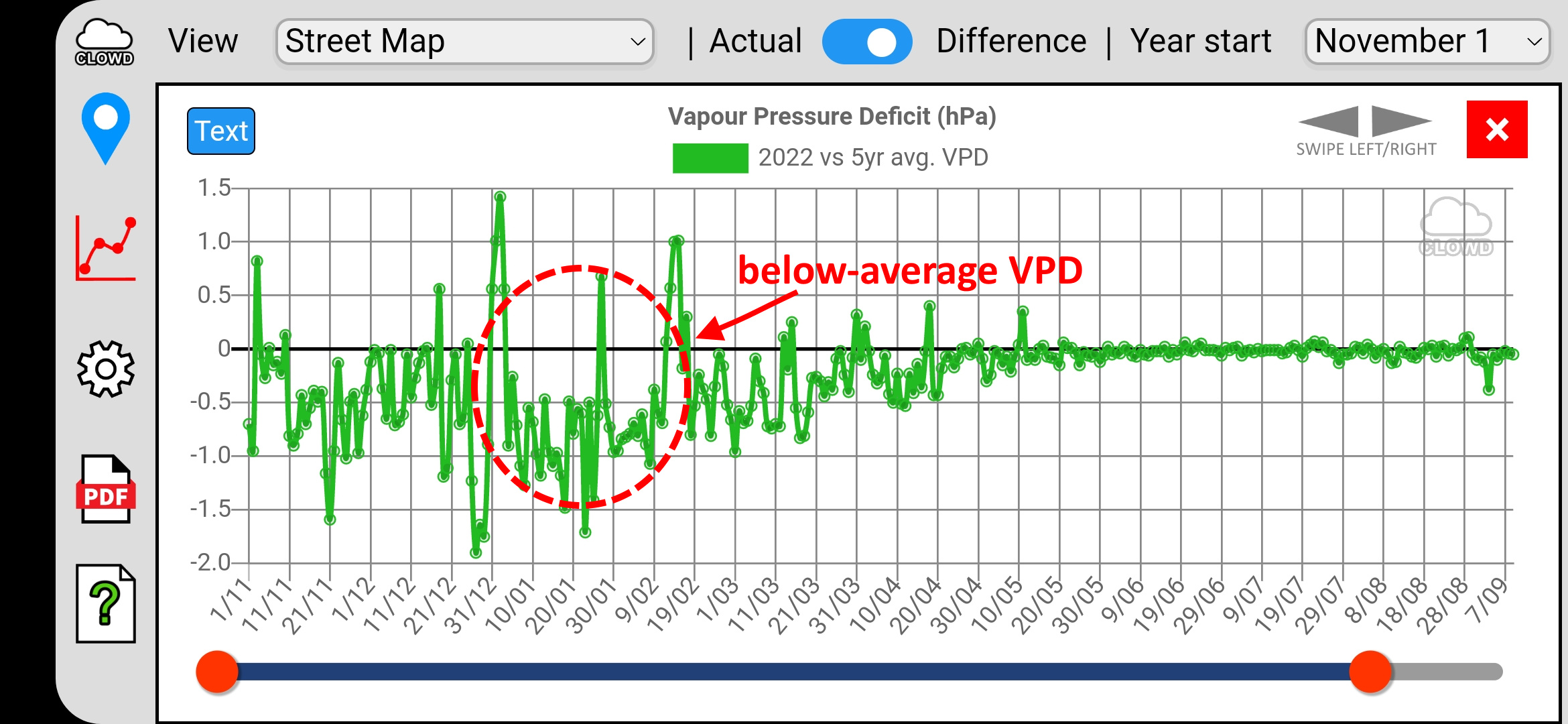}
\caption{Analysis using CLOWD's difference feature shows a period of below-average VPD during the key rice growth stage between panicle initiation and grain maturity at a selected rice-growing location.}
\label{fig:clowd10}
\end{figure}

The CLOWD framework does not provide the traditional DST response of telling a farmer or user when to perform a task or which task to perform. Rather, it is designed to enable the user to quickly identify weather patterns that may affect or explain certain crop outcomes. This section will now present an example of knowledge discovery achieved through CLOWD.\\
\indent During the 2021/22 rice growing season in southern New South Wales, it was noted that rice crops were maturing or reaching the `harvest' stage later than expected. One of the key growth parameters that affects the timing of the various rice phenology stages is vapour pressure deficit (VPD). A high VPD is seen to accelerate plant development by pulling nutrient-containing water through the plant, aiding that development, whereas a low VPD slows down that development process \citep{eca023}. Moreover, it has been reported that lower relative humidity and higher VPD can lead to higher head rice yield \citep{eca024}.\\
\indent As noted in Section \ref{sec:vpd}, CLOWD calculates its own VPD analysis, allowing comparison of the current growing season against any one of the previous five seasons or the five as an average. Using the smartphone version of CLOWD and the framework's `difference' feature, it was seen that at the chosen southern New South Wales location, the VPD was below the five-year average for almost the entirety of a six-week period between panicle initiation and grain maturity. This can be seen in Fig. \ref{fig:clowd10}. While VPD may not have been the only cause for late maturation, the fact this below-average period occurred is noteworthy nonetheless.\\
\indent The pictured view in Fig. \ref{fig:clowd10} was achieved by selecting the map location, swiping to the VPD chart and switching on the `difference' mode to show the chart data-value differences between the current season and the previous five-year average (the default comparison setting).

\begin{figure}
\centering
\includegraphics[width=1\columnwidth]{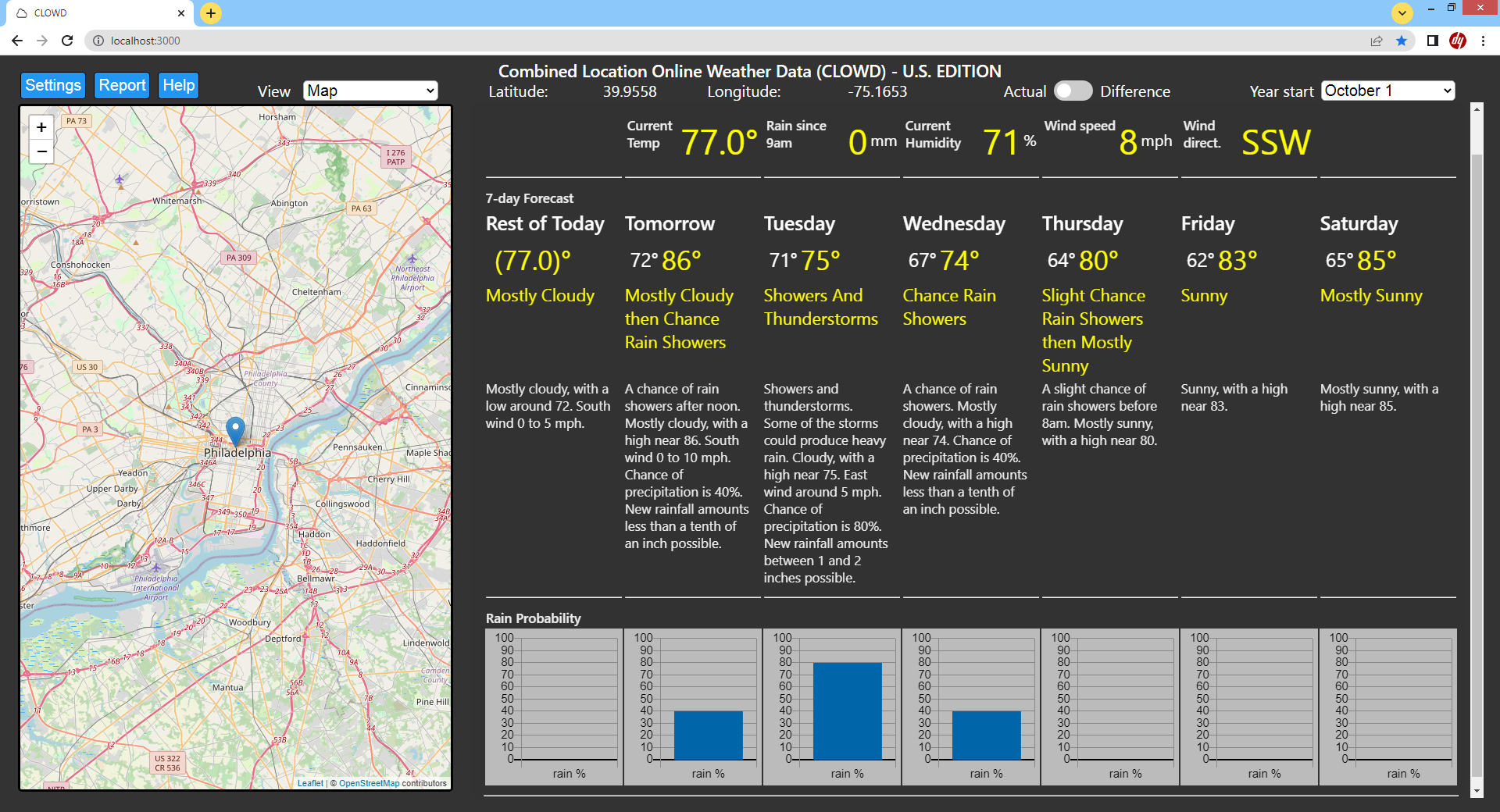}
\caption{An early prototype of a U.S. version of CLOWD showing location current and forecast weather using freely-available data from the U.S. National Oceanic and Atmosphere Administration (NOAA).}
\label{fig:clowd11}
\end{figure}

\section{Discussion and research opportunities}
While CLOWD has specific features aimed towards the agriculture industry (for example, the GDD charts), it is not limited to this industry. The new framework may have application in any field where fast analysis of recent or historical weather data is required, including bush/wildfires \citep{eca011} or flood mitigation \citep{eca012}. Further research is also being conducted to expand the CLOWD framework to include weather forecasting, allowing the addition of weather warnings, depending on the location and the warning-temperature settings selected. \\
\indent While the initial version of CLOWD is designed for Australia, the potential to transfer the framework to other geographic locations has been tested, with a working-prototype built around freely-accessible data made available by the U.S. National Oceanic and Atmospheric Administration (NOAA). This prototype, shown in Fig. \ref{fig:clowd11}, provides current weather conditions and a seven-day forecast (with percentage rain probability) for any location in the United States by clicking on the left-side map. Time permitting, our aim is to further complete this prototype and make it available for use.

\section{Conclusion}
Changing climate conditions are continuing to affect agricultural systems globally. This underlines the need for farmers and growers to have access to short- and long-term weather data to assist them in making the best decisions to achieve optimum outcomes. However, previous research has noted that despite the ubiquity of smartphone technology, the use of these `decision support tools' (DST) has not been universal. Moreover, the importance of a good user interface design to ensure a positive user experience has also been noted as a key factor in DST design moving forward. With these findings in mind, the proposed Combined Location Online Weather Data or `CLOWD' platform provides an easy-to-use method that is accessible from many computing devices and requires only a web browser. It features device-specific UIs and incorporates natural language generation (NLG) methods to simplify the understanding of the individual weather charts CLOWD provides. It is currently available as two separate prototypes - \url{https://clowd.csu.edu.au} for desktop/laptop computers, \url{https://clowds.csu.edu.au} for iOS and Android smartphones. While CLOWD was designed with Australian data in mind, a prototype implementation using U.S. weather data shows its potential for transfer to other locations, provided the required data is freely available.

\section*{Statements and Declarations}
This work was supported in part by funding from the Food Agility Co-operative Research Centre (CRC), funded under the Australian Commonwealth Government Co-operative Research Centre program, and in part by Charles Sturt University, AgriFutures Australia and Ricegrowers Ltd. One of us (AC) would like to acknowledge the receipt of a postgraduate research scholarship from Food Agility CRC.

\bibliography{CLOWD_v11_arxiv.bib}

\end{document}